\documentclass{PoS}
\usepackage{graphicx}
\usepackage{mwe}
\usepackage[mathscr]{euscript}
\usepackage{amsmath}
\usepackage{amsfonts}
\usepackage{physics}
\usepackage{latexsym,amsmath,amstext}
\usepackage{mathrsfs}
\usepackage[utf8x]{inputenc}
\usepackage{epsfig}
\usepackage{float}
\usepackage{slashed}
\usepackage{amssymb,fontenc,times,mathptmx,graphicx}

\title{Pion quasi parton distribution function on a fine lattice}

\ShortTitle{qPDF on fine lattice}
\author{
  \speaker{Charles Shugert}$^{1,4}$, Taku Izubuchi$^{1,2}$, Luchang Jin$^{2,3}$,
  Christos Kallidonis$^{4}$, Nikhil Karthik$^{1,4}$, Swagato Mukherjee$^{1}$,
  Peter Petreczky$^{1}$, Sergey Syritsyn$^{2,4}$\\
\llap{$^1$}Physics Department, Brookhaven National Laboratory, Upton, NY 11973, USA \\
\llap{$^2$}RIKEN-BNL Research Center, Brookhaven National Lab, Upton, NY, 11973, USA \\
\llap{$^3$}Physics Department, University of Connecticut, Storrs, Connecticut 06269-3046, USA \\
\llap{$^4$}Department of Physics and Astronomy, Stony Brook University, Stony Brook, NY 11794, USA
}

\abstract{
  We present a calculation of the bare quasi-PDF (qPDF) of the pion. We perform
  these calculations using the HotQCD HISQ gauge ensemble for our sea quarks along
  with a Wilson-Clover valence quark action. Our lattice size is $48^3\times64$,
  our lattice spacing is set at a = 0.06 fm, and our pion mass is tuned to 300 MeV.
  Utilizing momentum smearing techniques, we compute the bare qPDF boosted up to
  momentum 1.72 GeV. In addition we explore excited state contamination of the
  three-point correlator.
}

\FullConference{The 36th Annual International Symposium on Lattice Field Theory - LATTICE2018\\
		22-28 July, 2018\\
		Michigan State University, East Lansing, Michigan, USA.}

\begin{document}

\section{Introduction}
Deep inelastic scattering experiments can be used to probe the internal structure
of hadrons. The total cross-section in these interactions can be written as a
convolution of a partonic cross-section which can be computed perturbativly, and
a parton distribution function (PDF) defined as

\begin{equation}
  q(\xi) = \frac{1}{4\pi}\int d\xi^{-} e^{ixP^{+}\xi^{-}}
  \bra{N(P)}\bar\psi(\xi^-)\gamma^+W_L(\xi^-, 0)\psi(0)\ket{N(P)},
\end{equation}
where $\ket{N(P}$ is the hadron state,
$W_L(\xi^-, 0) = e^{ig\int_0^{\xi^-} d\xi^- A^+}$ is a straight Wilson Line on the
light-cone, and $\xi^{\pm} = (t \pm z)/\sqrt{2}$. PDFs, defined along the light-cone,
are inaccessible from direct lattice simulations due to the sign problem. Quasi-PDFs
have been proposed as an alternative method towards extracting PDFs, in which the
quark and anti-quark in the operator are only spatially separated \cite{Ji:2013dva}.

\begin{equation}
  q(x, P_z) = \frac{1}{4\pi}\int dz e^{-ixP^zz}
  \bra{N(P)}\bar\psi(z)\Gamma W_L(z, 0)\psi(0)\ket{N(P)},
\end{equation}
where $\Gamma$ is a gamma matrix is either $\gamma_z$ or $\gamma_t$. So long as the
hadron is highly boosted, matching the qPDF to the PDF can be done using large
momentum effective field theory \cite{Ji:2014gla}. 

Here we present calculations of the bare qPDF matrix element for a 300 MeV pion.
Our lattice size is $48^3\times64$ and our lattice spacing is 0.06 fm. We use a mixed
action using a HotQCD HISQ gauge ensemble \cite{Bazavov:2014pvz} and a Wilson-Clover
quark action with one level of HYP smearing \cite{Hasenfratz:2001tw}. In section two,
we study excited state effects on the two-point correlator under different smearing
schemes. We also examine the dispersion relation of the pion extracted from a
two-state fit of the two-point correlator to study lattice artifacts at large
momentum. In section three, we explore two methods, the summation method and a
two-state fit, that remove excited states from the three-point correlator. For all
matrix elements we used All-Mode Averaging (AMA) \cite{Shintani:2014vja} with 32
sloppy calculations to one exact solve for each configuration.

\section{Two-Point Function Analysis}
Here we present analysis of excited states and systematic error arising from the
two-point correlators. We first look at the effective mass plots at zero momentum
under different amounts of Wuppertal Smearing steps \cite{Gusken:1989ad}. To save
computation time, we switch to Coulomb gauge-fixed Gaussian smearing with a width set
to match the optimal Wuppertal smearing. We compute effective masses by solving the
following nonlinear equation
\begin{equation}
  C_{2pt}(t+1;P_z)/C_{2pt}(t;P_z) =
  \frac{\cosh{(M_{\text{eff}}(T/2-t+1;P_z))}}{\cosh{(M_{\text{eff}}(T/2-t;P_z))}},
\end{equation}
where $T$ is the length of the time-direction on our lattice, $t$ is a given
time-slice on the lattice, and $E$ is the effective mass. Figure 1. shows our
results. We can see that the effective mass plot reaches a plateau faster with 90
Wuppertal smearing than with only 40 steps. Furthermore, we find that appropriate
Gaussian smearing in Coulomb gauge has the same signal as 90 steps of Wuppertal
Smearing. For the width of the Gaussian we use is 0.31 fm.

As the signal of the two-point function deteriorates with larger momentum, we turn
to boosted sources, as prescribed in Ref. \cite{Bali:2016lva} for Coulomb
gauge-fixed Gaussian smearing. Here the valence quarks are boosted to momentum
$\vec{k} = \zeta \vec{P}$, where $\vec{P}$ is the momentum of our pion and $\zeta$ is
the momentum fraction of the quark. In Figure 2. we show effective mass plots of our
pion using boosted-smeared sources for different values of $\zeta$ at momentum 0.86
GeV, 1.29 GeV, and 1.72 GeV. At momentum 0.86 GeV $\zeta = 1.0$ yields data with
smaller error bars and smaller fluctuations about the expected mass from the
dispersion relation. At 1.29 GeV, $\zeta = 1.0$ is too large, and $\zeta = 0.67$
yields the best results. At 1.72 GeV, $\zeta = 0.5$ begins to fluctuate wildly early
on, whereas $\zeta = 0.75$ and $\zeta = 1.00$ are comparable to each other. Despite
this the error bars grow quickly and more statistics are needed at the momentum. For
momentum larger than 1 GeV however, we must omit data with $\zeta = 0$ as it becomes
too noisy to plot. And so, for large momentum calculations, it is necessary to have
large enough $\zeta$ in order to have a clean signal.
 
Next we study how well the pion with appropriate boosted smearing follows the expected
dispersion relation $E^2(P_z^2) = P_z^2 + m_{\pi}^2$. Here we fit our two-point
correlator data at fixed momentum to the following form:
\begin{equation}
  C_{\text{2pt}}(P_z, t)=\sum_{i=1}^{2}2A_i e^{-\frac{1}{2}E_iT}\cosh{(E_i(T/2 - t))},
\end{equation}
where $T$ is the time extent of the lattice, $t$ is the source-sink separation,
$A_i = |\bra{i}\ket{\pi}|^2$, and $E_i$ is the i'th energy eigenstate. Figure 2 shows
our results, and we find that there is no deviation from our extracted masses and the
expected dispersion relation of the pion. Table 1. shows $\Delta E_{2,1} = E_2 - E_1$
for momentum 0 GeV, 0.86 GeV, 1.29 GeV, and 1.72 GeV.
\begin{table}
  \centering
  \caption{Energy Gap vs $P_z$}
  \begin{tabular}{| c | c | c | c | c | c |}
    \hline
    $P_z$(N$_\text{cfg}$)&0 GeV (52)&0.86 GeV (168)&1.29 GeV (168)&1.72 GeV (168) \\
    \hline
    $\Delta E_{2,1}$ & 1.39(38) GeV & 1.26(04) GeV & 1.15(08) GeV & 1.32(36) GeV \\
    \hline
  \end{tabular}
\end{table}
%%       0 GeV       0.43 GeV    0.86 GeV    1.29 GeV    1.72 GeV
%% Ncfg  52          52          168         168         168
%% val   1.393406  , 1.2600787 , 1.25920444, 1.15333417, 1.3527734
%% err   0.37991418, 0.15064505, 0.04400372, 0.0819452 , 0.3626121
\begin{figure}
  \centering
  \begin{minipage}{0.45\textwidth}
    \centering
    \includegraphics[scale=0.475]{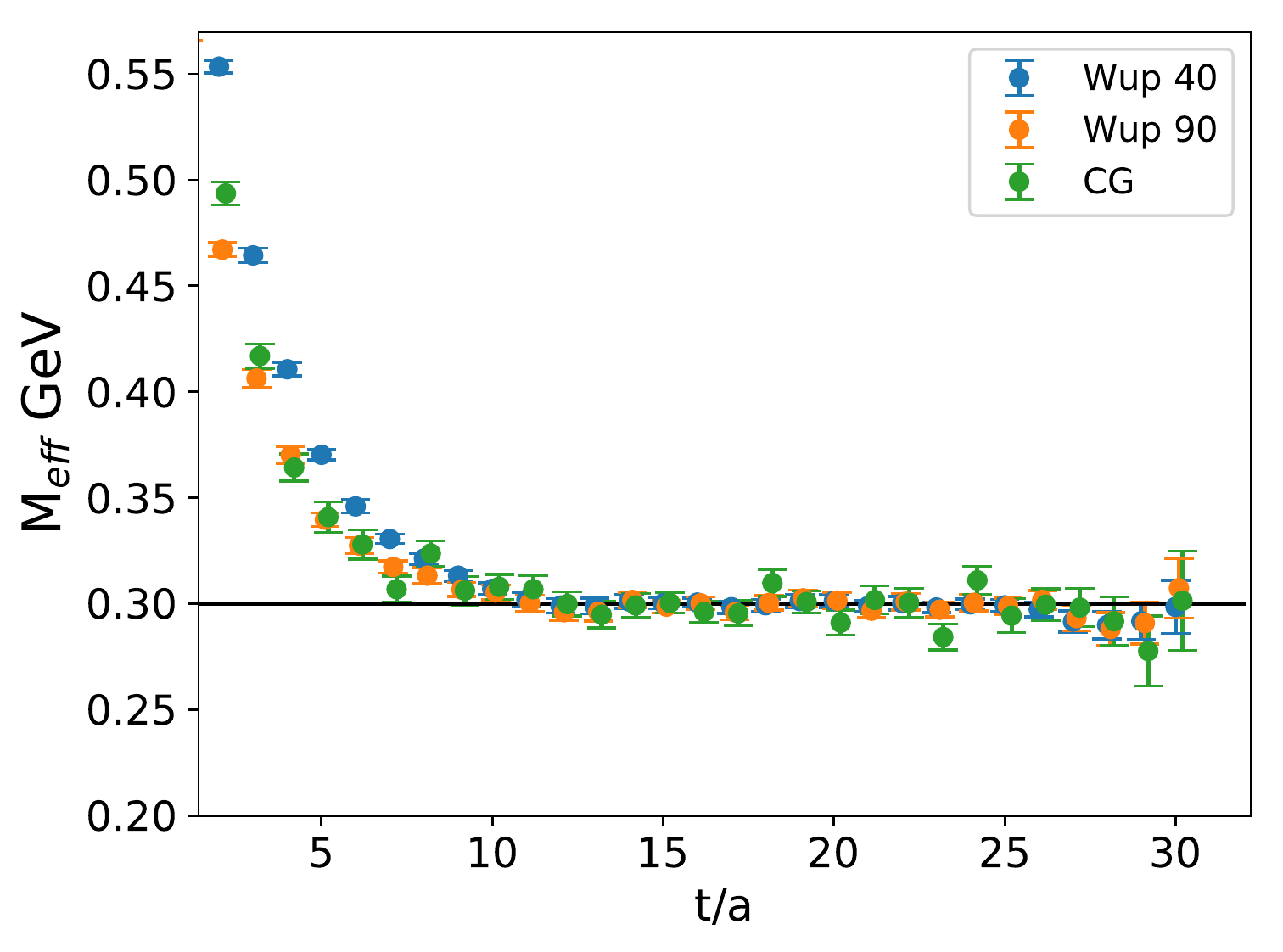}
    \caption{Effective Mass Plots at $P_z$ = 0 GeV using 50 configurations.
      Wup 40(90) correspond to 40(90) Wuppertal Smearing steps. CG
      stands for Coulomb gauge-fixed Gaussian smearing.}
  \end{minipage}\hfill
  \begin{minipage}{0.45\textwidth}
    \centering
      \includegraphics[scale=0.475]{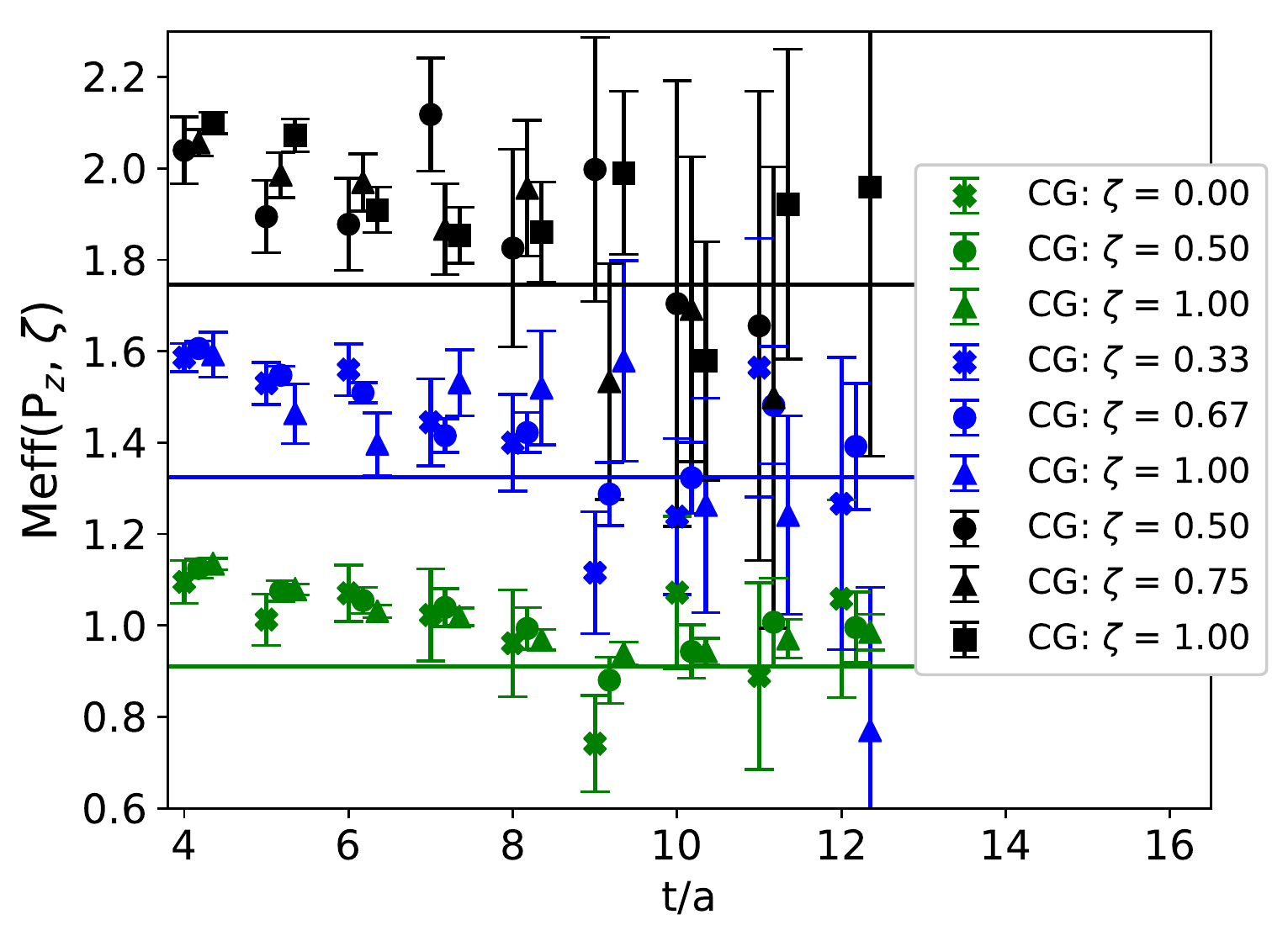}
      \caption{M$_{\text{eff}}$(P$_z$) vs different values of $\zeta$ with 50
        configurations, Green, blue, and black points correspond to momentum 0.86,
        1.29, and 1.72 GeV respectively.}
  \end{minipage}\hfill
\end{figure}

\begin{figure}
  \centering
  \begin{minipage}{0.45\textwidth}
    \centering
    \includegraphics[scale=0.45]{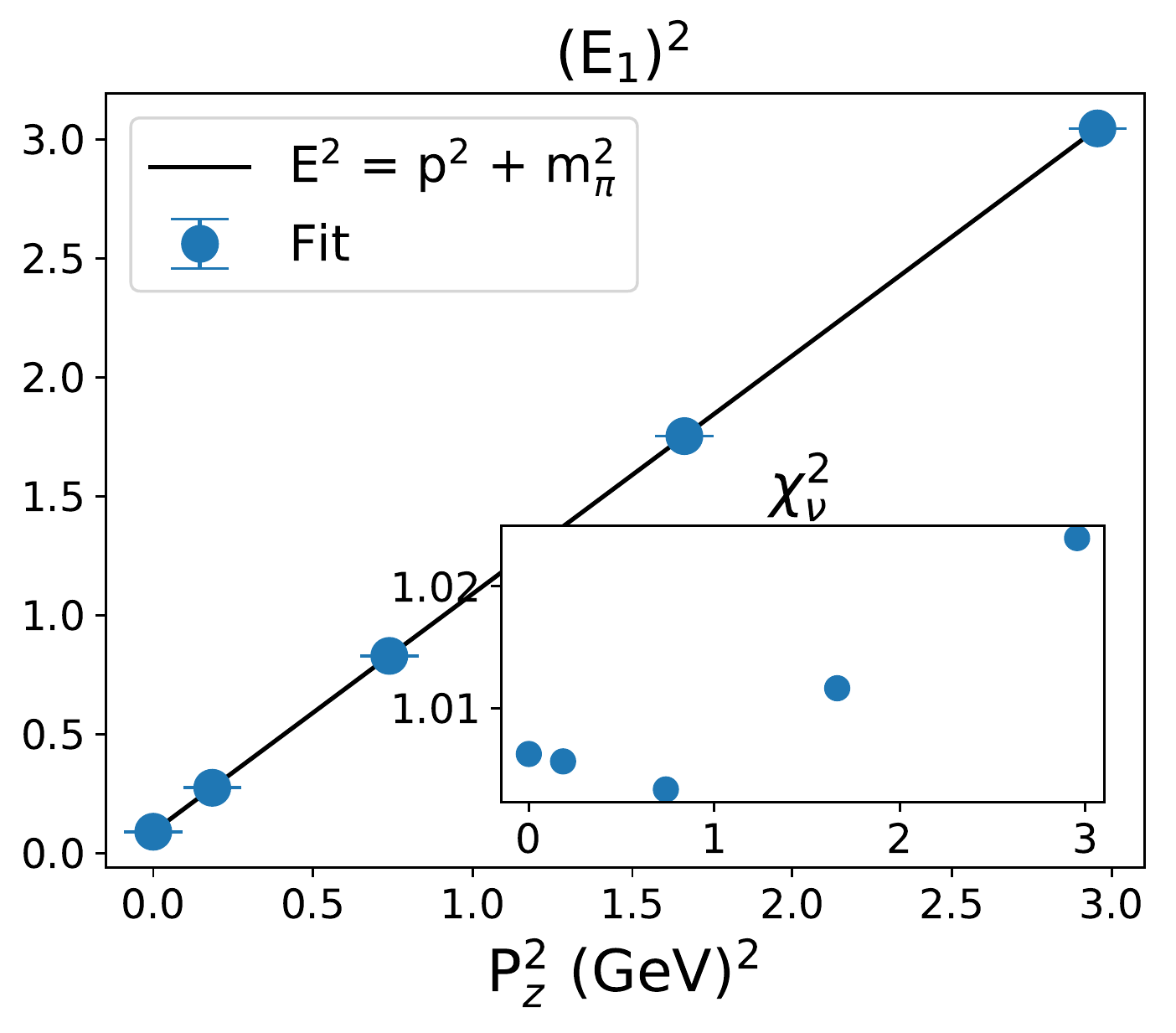}
    \caption{Ground-state energy squared vs $P_z^2$. Ground-state exctracted from
      fit to Eq (2.2).}
  \end{minipage}\hfill
  \begin{minipage}{0.45\textwidth}
    \centering
    \includegraphics[scale=0.4]{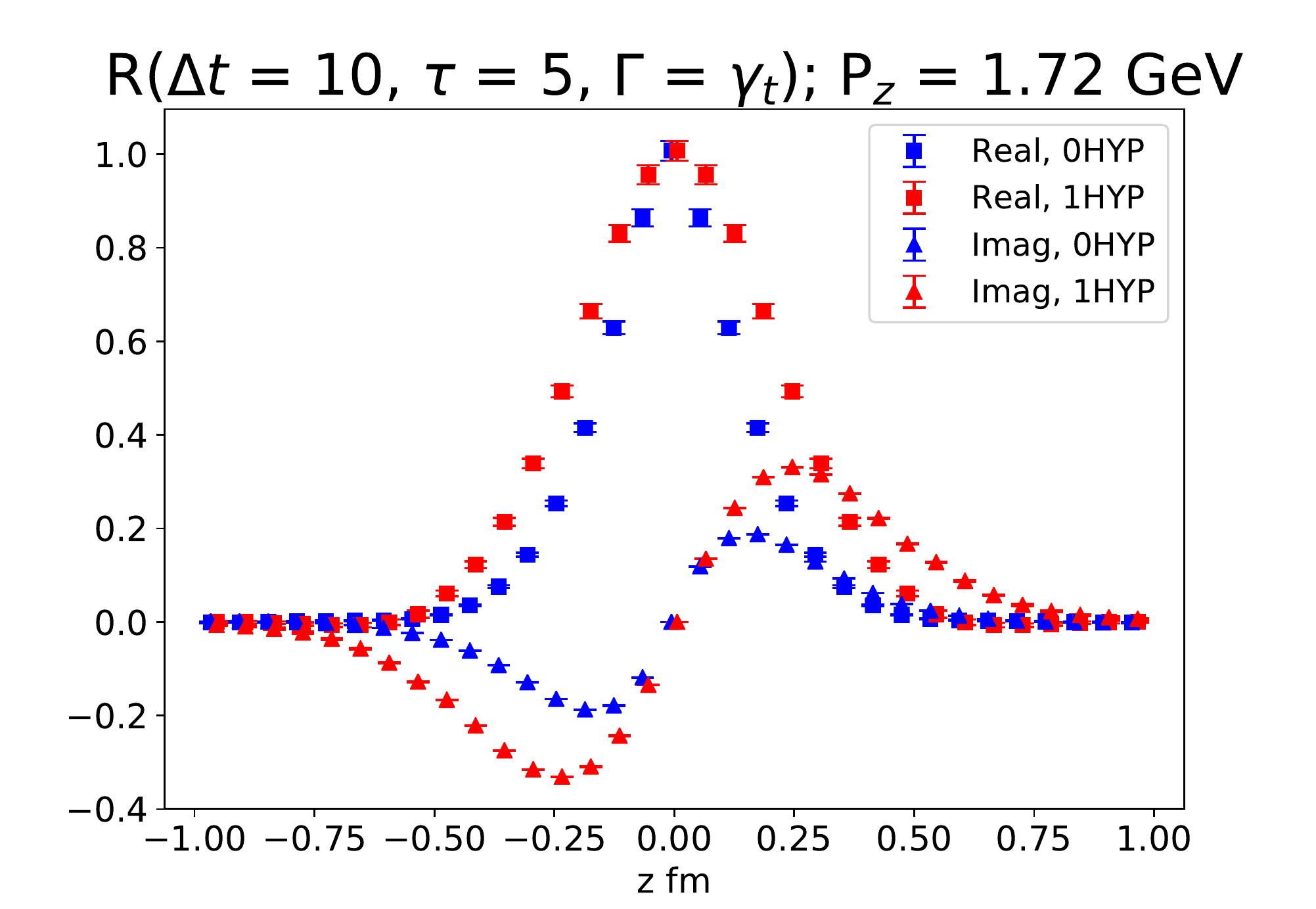}
    \caption{
      Ratio defined by Eq. (3.1) vs z under different applications of HYP smearing.
    }
  \end{minipage}\hfill
\end{figure}

\section{Three-Point Function Analysis}
To extract the ground state calculation of the three point function, we form the
ratio to remove the overlap factor between the source-operator and the pion state
\begin{equation}
  R(\Delta t, \tau z; \Gamma) = \frac
  {\langle\pi(\vec{p},\Delta t)\mathcal{O}_{\Gamma}(z,\tau)\bar{\pi(0)}\rangle}
  {\langle\pi(\vec{p},\Delta t)\bar{\pi(0)}\rangle}
  = \frac{\sum_{n,n'}A_nA_{n'}^*e^{-E_n\Delta t}e^{-(E_{n'}-E_n)\tau}\bra{n}\mathcal{O}_{\Gamma}(z)\ket{n'}}{\sum_m |A_{m}|^2e^{-E_m \Delta t}}
\end{equation}

where $A_n = \bra{\pi}\ket{n}$,
$\mathcal{O}_{\Gamma}(z, \tau) = \bar\psi(z,\tau)\Gamma W_L(z,0)\psi(0)$,
$\Delta t$ is the source-sink separation, and $\tau$ is the operator insertion
time-slice, such that $0 < \tau < \Delta t$. Expanding this to the second energy
eigenstate yields
\begin{equation}
  R(\Delta t, \tau, z; \Gamma) \sim
  \frac{\mathscr{M}(z) + \mathscr{A}(z)e^{-\Delta E_{2,1}\tau} +
    \mathscr{A}^\dagger(z)e^{-\Delta E_{2,1}(\Delta t-\tau)} +
    \mathscr{B}(z)e^{-\Delta E_{2,1}\Delta t} + ...}
  {1 + \mathscr{C}e^{-\Delta E_{2, 1}\Delta t} + ...}.
\end{equation}
Here, $\mathscr{M}(z) = \bra{1}\mathcal{O}_{\Gamma}(z)\ket{1}$ is the desired
quantity,
$\mathscr{A}(z) = \frac{A_1A_2^*}{|A_1|^2}\bra{1}\mathcal{O}_{\Gamma}(z)\ket{2}$,
$\mathscr{B}(z) = \mathscr{C}\bra{2}\mathcal{O}_{\Gamma}(z)\ket{2}$,
and $\mathscr{C} = \frac{|A_2|^2}{|A_1|^2}$, a term suppressed for large
$\Delta E_{2,1}\Delta t$.

In Figure 4 we display a comparison of our bare quasi-PDF matrix element when we
apply 0 and 1 HYP smearing operations onto the Wilson Line of our operator
\cite{Hasenfratz:2001tw}. We do so to see whether or not lattice artifacts due to
short distance fluctuations of the gauge field can be smeared out with HYP smearing.
Here $\Delta t$ = 10, and the operator insertion $\tau = \Delta t/2$. What we
see is that the matrix element becomes wider and larger with respect to the length
of the Wilson Line. This is expected, as the self energy divergence from the Wilson
Line decreases with HYP smearing. However, the statistical error in the matrix element
does not improve much with HYP smearing.

The data in Figure 5 showcases data with 1 HYP-Smeared Wilson Line. Points on the same
z-value are shifted horizontally for better visability. In addition all z values
greater than 0.48 fm in magnitude are computed with 52 configurations, whereas all
smaller z values are computed using 168 configurations. Here we plot the ratio vs the
length of the Wilson Line with respect to $\Delta t$ = 8, 10, and 12, with
$\tau = \Delta t/2$. On one hand the error in smaller $\Delta t$ are very small and
increases with increasing $\Delta t$. On the other hand, excited state contamination
dies off exponentially with larger source-sink separation. And so, to extract the
ground-state quasi-PDF matrix element, we employ two fitting procedures used in
references \cite{Abdel-Rehim:2015owa, Maiani:1987by}.

\begin{figure}
  \centering
  \includegraphics[width=\textwidth]{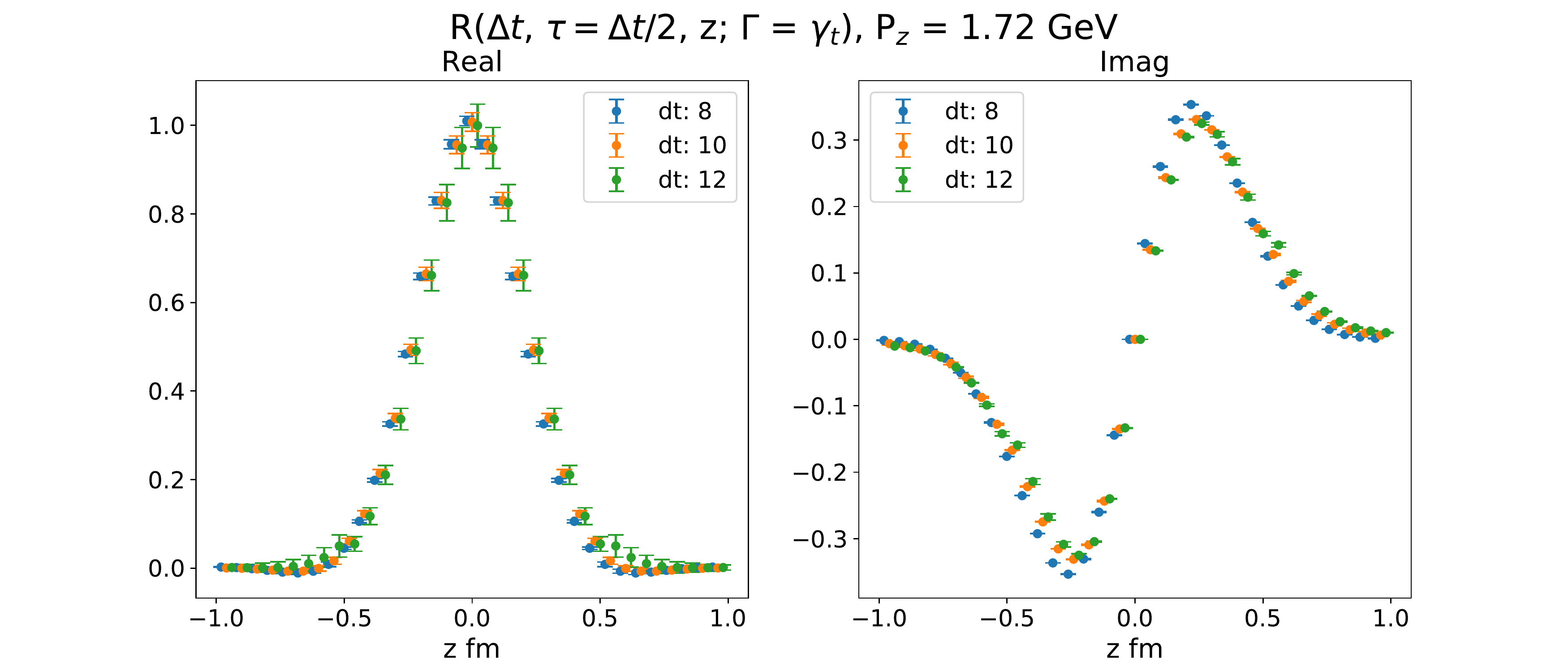}
  \caption{Ratio defined by Eq. (3.1) vs z at $\Delta t$ 8 (blue), 10 (orange),
    and 12 (green). $\tau = \Delta t/2$.}
\end{figure}
The first method employed is the summation method \cite{Maiani:1987by}. Here one
takes the ratio and sums over all operator insertions minus a certain number of
end points. Here we sum the ratio excluding $\tau > |\tau_o|$. The summed
ratio takes the following form
\begin{equation}
  R_{\text{sum}}(\Delta t, z; \Gamma) =
  \sum_{\tau=\tau_o}^{\Delta t - \tau_o}R(\Delta t, \tau, z; \Gamma) \sim
  (\mathscr{M} + \mathscr{B}e^{-\Delta E_{2,1} \Delta t})(\Delta t - 2\tau_o) + C,
\end{equation}
where $\tau_o$ represents the time slices truncated from the sum. For each source-sink
separation, we compute the summed ratio and plot it with respect to
$\Delta t - 2\tau_o$. From there we perform a linear fit to extract the slope,
which gives an estimate of $\mathscr{M}$ for large enough $\Delta E_{2,1}\Delta t$.
In figures 6 and 7, we plot the results of the summation method for the ratio
evaluated at z = 0 fm and z = 0.24 fm using $\tau_o = 1$ and $\tau_o = 2$ to determine
which gave the more precise result. In both cases $\tau_o = 2$ gives the most precise
result. As this method does not remove some of the systematic error from excited
states, we also try a second method.
\begin{figure}
  \centering
  \begin{minipage}{0.4\textwidth}
    \centering
    \includegraphics[scale=0.45]{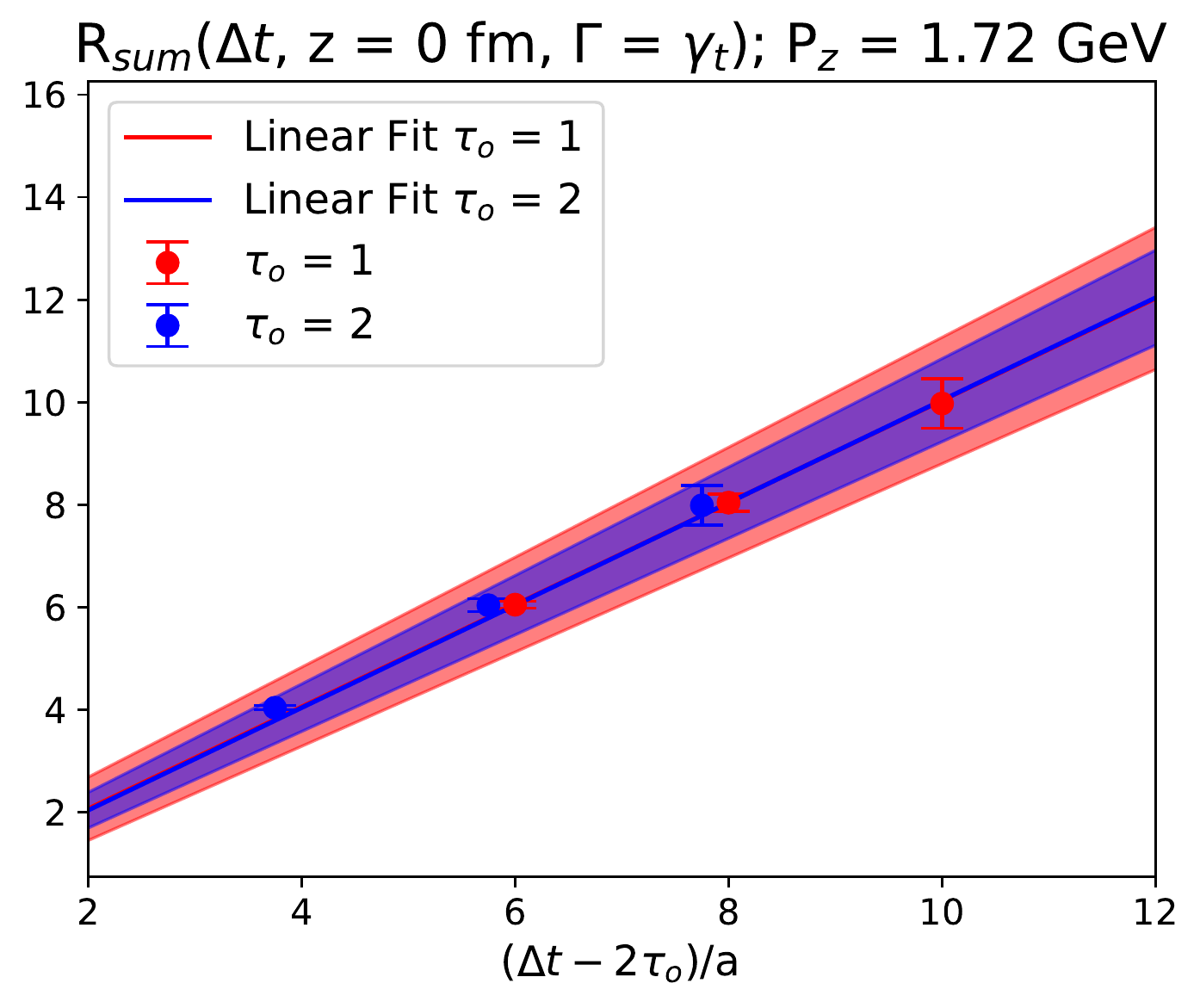}
    \caption{Summation method results for z = 0 fm. Red and blue data are for
      results with $\tau_o$ = 1 and $\tau_o$ = 2 respectively. The line and
      bands about the line correspond to the results of the linear fit and
      the error bands to that fit result.}
  \end{minipage}\hfill
  \begin{minipage}{0.4\textwidth}
    \centering
    \includegraphics[scale=0.45]{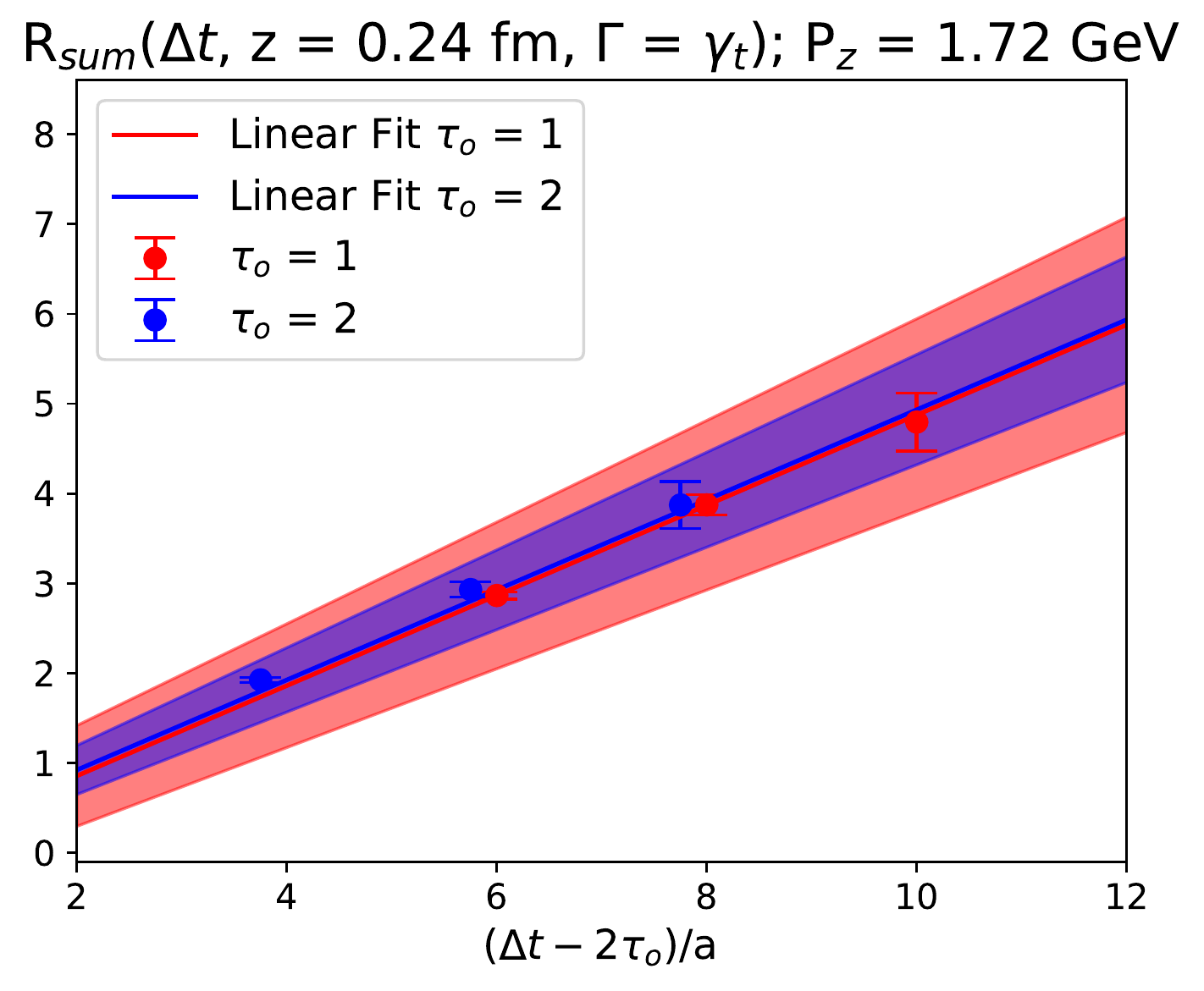}
    \caption{Summation method results for z = 0.24 fm. Red and blue data are
      for results with $\tau_o$ = 1 and $\tau_o$ = 2 respectively. The line and
      bands about the line correspond to the results of the linear fit and
      the error bands to that fit result.}
  \end{minipage}\hfill
\end{figure}
The second method relies on a simultaneous fit of both the two-point and three-point
correlator, which we refer to as the two-state fit \cite{Abdel-Rehim:2015owa}. Using
$\Delta E_{2,1}$ obtained from fitting the two-point correlator, we use all operator
insertions excluding end points as well as all available source-sink separation data
to fit $\mathscr{M}$, $\mathscr{A}$, and $\mathscr{B}$ in equation (3.2).

On Figure 8 we plotted the results of the two-state fit and the summation method along
with the ratio at $\Delta t$ = 10 with the $\tau = \Delta t/2$. Again, the data
contains a Wilson Line under one application of HYP smearing. All z values greater
than 0.48 fm in magnitude are computed with 52 configurations and the rest with 168
configurations, with points on the same z-value shifted horizontally for better
visability. In addition all the points on the same z-value are shifted horizontally
for better visability. We see that almost consistently the results from the summation
method give larger error. Data beyond 0.48 fm yields larger error, consistent with the
smaller configurations used compared to data with z < 0.48 fm. The imaginary part of
the summation method results does differ somewhat to the imaginary part of the
two-state fit, meaning that excited states have some effects in the imaginary part. 
\begin{figure}
  \centering
  \includegraphics[width=\textwidth]{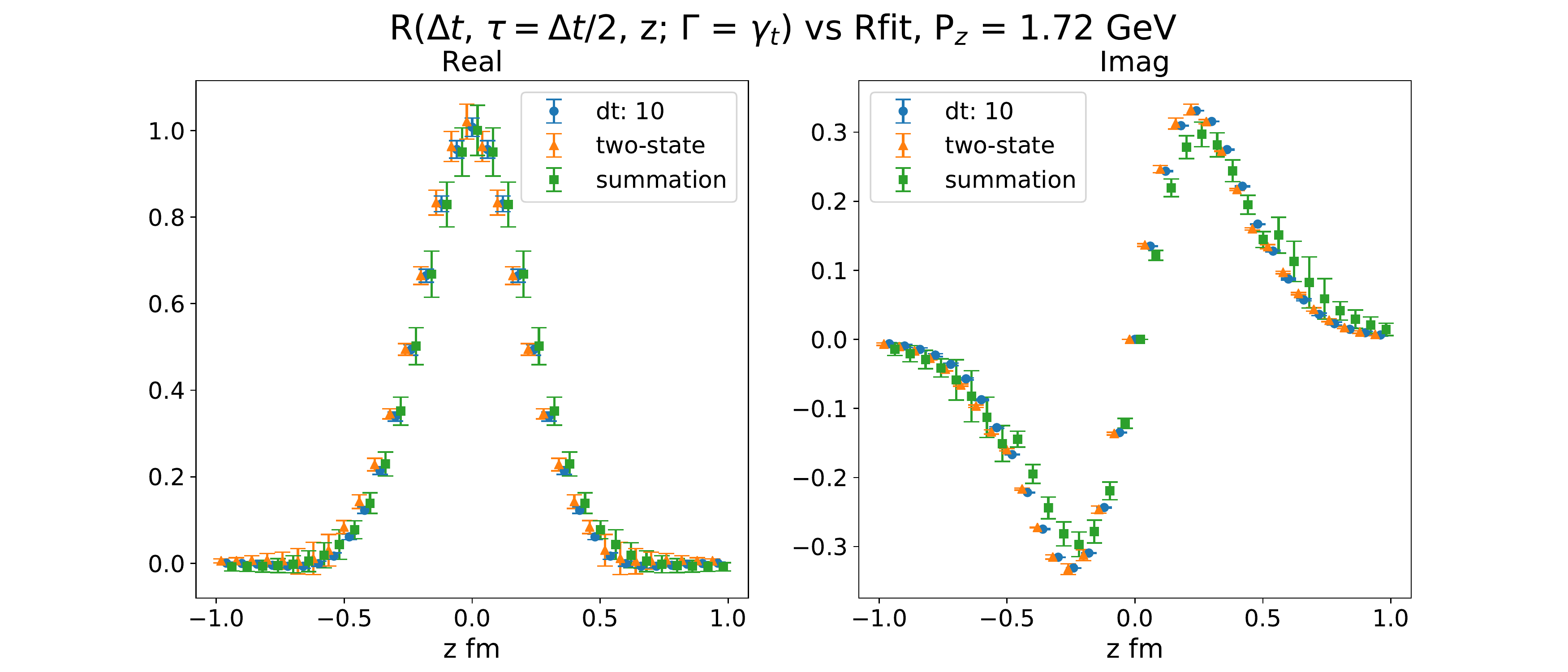}
  \caption{Ratio defined by Eq. (3.1) vs z
    at $\Delta t$ = 10 $\tau = \Delta t/2$ (blue),
    extracted ground state matrix element from a two-state fit (orange), and
    extracted ground state matrix element from the summation method.}
\end{figure}
\section{Conclusion}
We have presented the calculation of the bare quasi-PDF matrix element at momentum
1.72 GeV for a lattice spacing 0.06 fm using a mixed HISQ-sea and a 1 HYP-smeared
valence Wilson-Clover fermion action for a 300 MeV pion. In order to maximally project
our pion field onto the ground state, we studied different smearing techniques. We
found optimal Gaussian sources to project onto the pion ground state, and have
increased the signal of our data at large momentum using boosted sources. We also
studied the ground state energy of the pion, extracted from a double exponential fit
up to momentum 1.72 GeV, and found that agrees with the expected dispersion relation
$E^2 = p^2 + m_{\pi}^2$. Regarding the quasi-PDF matrix element, we looked at the
HYP-smearing dependence. Furthermore we explored excited state removal of our
three-point correlator using a simultaneous fit of the two and three-point correlator
and by applying the summation method. 

\acknowledgments
C.S, T.I, S.M, P.P. and N.K. acknowledge support by the U.S. Department of Energy
under contract No. DE-SC0012704, BNL LDRD project No. 16-37 and Scientific Discovery
through Advance Computing (SCIDAC) award ”Computing the Properties of Matter with
Leadership Computing Resources”. The computations were carried out using USQCD
facilities at JLab and BNL under a USQCD type-A project. This research also used an
award of computer time provided by the INCITE program at the Oak Ridge Leadership
Computing Facility, which is a DOE Office of Science User Facility supported under
Contract DE-AC05-00OR22725.

%% C.S, T.I., S.M., P.P. and N.K acknowledge support by the U.S. Department of Energy
%% under contract No. DE-SC0012704, BNL LDRD project No. 16-37. The computations were
%% carried out using USQCD facilities under a USQCD type-A project. The work was
%% supported by the U.S. Department of Energy, Office of Nuclear Physics through the
%% Contract No. DE-SC001270 and Scientific Discovery through Advance Computing (SCIDAC)
%% award ”Computing the Properties of Matter with Leadership Computing Resources”.
%% Computations were carried out using USQCD resources at JLab and BNL. This research
%% also used an award of computer time provided by the INCITE program at the Oak Ridge
%% Leadership Computing Facility, which is a DOE Office of Science User Facility
%% supported under Contract DE-AC05-00OR22725. 

\bibliographystyle{JHEP}
\bibliography{posbib.bib}

%% \begin{thebibliography}{99}
%% \bibitem{...}
%% ....

%% \end{thebibliography}

\end{document}